\newif\ifsubmit
\submittrue

\newif\ifarxiv
\arxivtrue

\newif\ifdraft
\draftfalse

\ifdraft
\documentclass[draft]{\ifsubmit\else ../../sty/llncs/\fi llncs}
\else
\documentclass{\ifsubmit\else ../../sty/llncs/\fi llncs}
\fi

\newif\iffinal
\finaltrue

\usepackage[T1]{fontenc}
\usepackage[utf8]{inputenc}
\usepackage{lmodern}
\usepackage[scaled=.8]{beramono}
\usepackage{latexsym}
\usepackage{textcomp}

\usepackage{amssymb}
\usepackage[final]{graphicx}
\usepackage{caption}
\usepackage{listings}
\usepackage{booktabs}
\usepackage{mathtools}
\usepackage{tabularx}

\usepackage{tikz}
\ifsubmit
\usetikzlibrary{arrows,calc,positioning,shapes}
\else
\usetikzlibrary{arrows.meta,calc,positioning,shapes}
\fi

\usepackage[ngerman,british]{babel}
\usepackage[inline]{enumitem}
\usepackage[babel]{csquotes}
\MakeAutoQuote{“}{”}
\MakeAutoQuote*{‘}{’}
\usepackage{soul}
\setul{.25ex}{}

\usepackage[obeyDraft,textsize=tiny]{todonotes}

\usepackage[backend=bibtex,hyperref=auto,
firstinits=true,style=numeric,
urldate=iso8601,isbn=false,doi=false]{biblatex}
\ifsubmit
\else
\fi
\DeclareNameAlias{author}{last-first}
\DeclareNameAlias{editor}{last-first}
\DeclareNameAlias{translator}{last-first}
\DeclareFieldFormat{labelnumberwidth}{#1.}
\DeclareFieldFormat{title}{#1}
\DeclareFieldFormat
  [article,inbook,incollection,inproceedings,patent,thesis,unpublished]
  {title}{#1}
\DeclareFieldFormat{journaltitle}{#1}
\newbibmacro*{volume+number+eid}{%
  \printfield{volume}%
  \iffieldundef{number}
    {}
    {(%
      \printfield{number}%
     )%
    }%
  \setunit{\addcomma\space}%
  \printfield{eid}}
\DeclareFieldFormat{url}{\url{#1}}

\usepackage{hyperref}

\begin{document}

\title{Semantic Publishing Challenge –\newline Assessing the Quality of Scientific Output\ifarxiv\thanks{The final publication is available at \texttt{link.springer.com}}\fi}
\author{Christoph Lange\inst{1}
\and Angelo Di Iorio\inst{2}}
\institute{%
University of Bonn \& Fraunhofer IAIS, Germany
\email{math.semantic.web@gmail.com}
\and Università di Bologna, Italy
\email{diiorio@cs.unibo.it}
}

\maketitle

\begin{abstract}
Linked Open Datasets about scholarly publications enable the development and integration of sophisticated end-user services; however, richer datasets are still needed. 
The first goal of this Challenge was to investigate novel approaches to obtain such semantic data.
In particular, we were seeking methods and tools to extract information from scholarly publications, to publish it as LOD, and to use queries over this LOD to assess quality.
This year we focused on the quality of workshop proceedings, and of journal articles w.r.t.\ their citation network. 
A third, open task, asked to showcase how such semantic data could be exploited and how Semantic Web technologies could help in this emerging context.
\end{abstract}

\section{Introduction: Scholarly Publishing and the Semantic Web}
\label{sec:schol-publ-semant}

Scholarly publishing is increasingly driven by a new wave of applications that better support researchers in disseminating, exploiting and evaluating their results.
The huge potential of publishing scientific papers enriched with semantic information has been proved, e.g., by Elsevier's Grand Challenges of 2009 and 2011~\cite{Elsevier:GrandChallenge09,Elsevier:EPC11} and by the yearly SePublica~\cite{SEPUBLICA14} and Linked Science workshop series~\cite{LinkedScience2013}\footnote{\url{http://linkedscience.org/category/workshop/}}, both taking place for the fourth time in 2014.
The semantic publishing community believes that 
semantics will help to improve the way users access, share, exploit and evaluate research results, and it will help to advance services such as search, expert finding, or visualisation, and even further applications not yet envisioned.
Semantic Web technologies 
play a central role in this context, as they can help publishers to make scientific results available in an open format the whole research community can benefit from.
New ways of publishing scientific results, as presented at the events mentioned above, include:
\begin{itemize}
\item machine-comprehensible experimental data,
\item linking machine-comprehensible datasets to research papers,
\item machine-comprehensible representations of scientific methods and models,
\item alternative publication channels (e.g.\ social networks and micro-publications),
\item alternative metrics for scientific impact (“altmetrics”~\cite{altmetrics-manifesto}), e.g., taking into account the scientist's social network, user-generated micro-content such as discussion post, and recommendations.
\end{itemize}
Scientific data published using Semantic Web technology not only solves isolated problems, but generates further value in that datasets can be shared, linked to each other, and reasoned on.

Section~\ref{sec:chall-defin-publ} explains how we developed the definition of this year's Semantic Publishing Challenge, Section~\ref{sec:extraction-tasks} explains the evaluation procedure for the two information extraction tasks, Sections~\ref{sec:task1} to \ref{sec:task3} explain the definitions and outcomes of the three tasks in detail, and Section~\ref{sec:over-less-learn} discusses overall sessions
learnt.

\section{The Challenge of Defining a Publishing Challenge}
\label{sec:chall-defin-publ}

In the other two challenges in the ESWC Semantic Web Evaluation Challenges track, it seemed straightforward to objectively measure the performance of a solution, as suitable, curated datasets existed – the Blitzer dataset for the Concept-Level Sentiment Analysis Challenge~\cite{RC:CLSA14} – or were relatively straightforward to obtain – the DBbook dataset for the Linked Open Data-enabled Recommender Systems Challenge~\cite{NCO:RecSys14}.

Existing datasets 
on
scholarly publishing mainly contain basic bibliographical metadata (such as DBLP~\cite{DBLP}), or research data specific to one scientific domain, as can, e.g., be seen from the “life science” section of the LOD Cloud~\cite{State-LOD-Cloud}.
In preparing the challenge, we judged that basic bibliographical author/title/year metadata did not have a sufficiently challenging semantics, whereas advanced publishing applications could be built on top of richer semantic data;
we also judged that existing research datasets were too domain-specific to design relevant and feasible challenge tasks around them.
We concluded that the semantic publishing community had so far lacked datasets adequate for evaluation challenges and therefore designed our \emph{first} challenge\footnote{\url{http://challenges.2014.eswc-conferences.org/index.php/SemPub/}} to \emph{produce, by information extraction, an initial collection of data that would be useful for future challenges and that the community can experiment on}.
This data collection should, of course, be produced in an objectively measurable way.

Technically, we focused on producing data \emph{going beyond basic bibliographical metadata} in terms of structure and complexity
, and in that some of the information
covered
would not be available from existing structured databases, but only by full-text analysis;
consider, e.g., authors' affiliations, or references to funding bodies.
As “producing data to base future challenges on” is not an appealing objective in itself, we identified \emph{quality} as a key concern of relevance to the whole scientific community: \emph{how can one assess the quality of scientific production} by automated data analysis?
This time, we focused on analysing (extended) metadata of publications, not yet on research data or on linked data representations of the full 
text of publications%
.
The objective of Task~1 of 3 was to assess the quality of workshops by computing metrics from data extracted from their proceedings, also considering information about persons and events.
The objective of Task~2 was to assess the quality of journal articles by characterising citations and identifying, e.g., their context, their function and their position in the citing papers.

After calling for submissions to Tasks~1 and 2, we received feedback from the community that mere information extraction, even if motivated by quality assessment, was not the most exciting task related to the future of scholarly publishing, as it assumed a traditional publishing model where results are mainly disseminated through papers, and the quality of scientific production is assessed from these papers.
We therefore added a third, \emph{open task}.

\section{Common Procedures for the Extraction Tasks}
\label{sec:extraction-tasks}

The extraction tasks~1 and 2 followed a common procedure similar to the other evaluation challenges:
\begin{enumerate}
\item For each task, we initially published a \emph{training dataset} (TD) on which the participants could test and train their extraction tools.
\item We specified the basic structure of the linked data to be extracted from these source data, without prescribing a vocabulary.
\item We provided natural language queries and their expected results on TD.
\item A few days before the submission deadline, we published an \emph{evaluation dataset} (ED), a superset of TD, which was the input for the final evaluation.
\item We asked the participants to submit their extracted linked data (under an open license to permit 
reuse%
), SPARQL implementations of 
the queries%
, as well as their extraction tools, as we reserved the right to inspect them.
\item We awarded prizes for the best-performing
(w.r.t.\ precision/recall)
and for the most innovative approach
(to be determined by an expert jury).
\item Both before and after the submission we maintained transparency.
Prospective participants were invited to ask questions, e.g. about the expected query results, which we answered publicly.
After the evaluation, we made the scores and the gold standard (see below) available to the participants.
\end{enumerate}

\begin{figure}
  \centering
  \begin{minipage}{.53\textwidth}
    \begin{tikzpicture}
      \ifsubmit
\begin{scope}[
  dataset/.style={draw,cylinder,shape border rotate=90,aspect=.25,minimum width=3em,minimum height=9ex,font={\scriptsize}},
  file/.style={draw,rectangle,minimum width=3em,minimum height=8ex,font={\scriptsize}},
  process/.style={->,>=stealth},
  process-label/.style={align=center,font={\tiny}},
  align=center,
  node distance=.35cm]
  \node[dataset] (source) {Source\\ data};
  \node[dataset,right=of source] (lod) {Linked\\ data\\ (RDF)};
  \node[file,right=of lod] (result) {Query\\ result\\ (CSV)};
  \node[file,below=.5cm of result] (gold) {Gold\\ standard\\ (CSV)};
  \coordinate (mid) at ($(result.south)!.5!(gold.north)$);
  \node[file,right=1.5cm of mid] (report) {Report\\ (HTML)};
  \draw[process,bend left] (source.north) to node[process-label,above] {extraction\\ tool} (lod.north);
  \draw[process,bend left] (lod.north) to node[process-label,above] {query} (result.north);
  \node at (mid) [process-label,anchor=west] {evaluation\\ script};
  \draw[process] (result) to (report);
  \draw[process] (gold) to (report);
\end{scope}

      \fi
    \end{tikzpicture}
    \captionof{figure}{Precision/recall evaluation}
    \label{fig:eval}
  \end{minipage}%
  \begin{minipage}{.44\textwidth}
    \includegraphics[width=\textwidth]{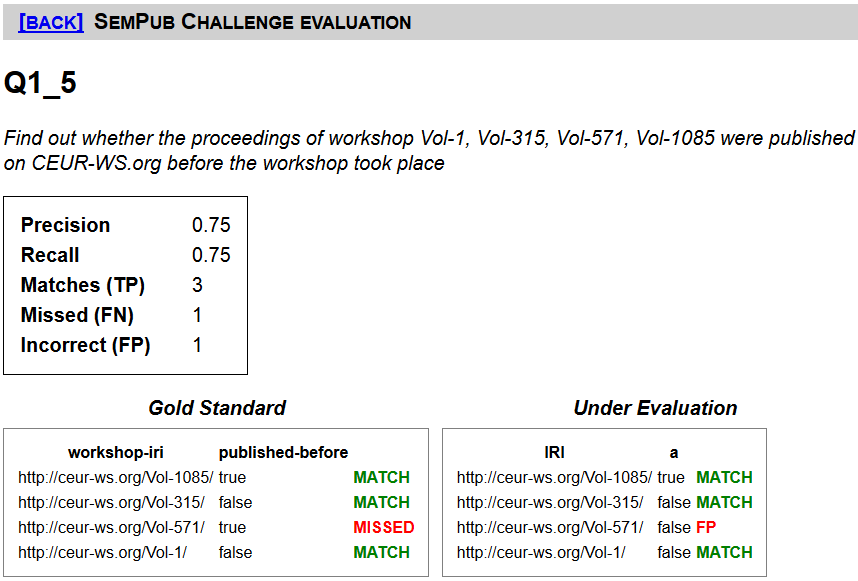}
    \captionof{figure}{Report for one query}
    \label{fig:report}
  \end{minipage}
\end{figure}

The given queries contained placeholders, e.g. 
“all authors of the paper titled $T$”.
For training, we specified the results expected after substituting certain values from TD for the variables.
We evaluated by substituting further values, mostly values that were only available in ED.
We had intentionally chosen easy as well as challenging queries, all weighted equally, to help participants get started, without sacrificing our ability to clearly distinguish the best-performing approach.
The 
evaluation was automated with a collection of PHP scripts: they compared a CSV form of the results of the participants' SPARQL queries over their data against a gold standard of expected results
, and compiled 
a
report with precision/recall measures and a list of false positives and false negatives (see Figures~\ref{fig:eval} and \ref{fig:report}).

\section{Task 1: Extraction and Assessment of Workshop Proceedings Information}
\label{sec:task1}

\subsection{Motivation and Objectives}
\label{sec:objectives1}

Common questions related to the quality of a scientific workshop or conference include whether a researcher should submit a paper to it, whether a researcher should accept an invitation to its programme committee, whether a publisher should publish its proceedings, and whether a company should sponsor it~\cite{BrylEtAl:SePublica2014}.
Moreover, knowing the quality of a scientific event helps to assess the quality of papers that have been accepted there.
Quality indicators include
\footnote{Some of these indicators have been suggested by Manfred Jeusfeld, the founder and publisher of CEUR-WS.org.}
a long history and growth over time,
attracting high-quality sub-events,
a high ratio of contributed over invited papers
(unless there are high-profile invited speakers)
,
a low ratio of submissions (co-)authored by the event's chairs (and thus a high diversity in schools of thought),
a fast publication turnaround (proceedings published quickly after, or even before the workshop – giving an impression of professional organisation).

Producing data that would help to answer such questions was the first objective of Task~1, from the perspective of the Semantic Publishing Challenge.
There was a second motivation, owed to Christoph Lange's role of technical editor of CEUR-WS.org.
CEUR-WS.org is an open access publishing website which enjoys great popularity among the organisers of computer science workshops and has published more than 1,200 proceedings volumes since 1995.
CEUR-WS.org has recently started discussing innovations\footnote{\url{http://ceurws.wordpress.com}}, and a linked data representation of the workshop proceedings would certainly facilitate the implementation of innovative services.
As the CEUR-WS.org volumes also provide sufficient information for answering quality-related questions such as those listed above, we chose them as the data source for Task~1.
Note that DBLP also covers most of the workshops published with CEUR-WS.org, but it does not cover certain quality-related information, such as what series a workshop is part of, the affiliations of the editors, the exact dates of a workshop and the publication of its proceedings, and a distinction between invited and contributed papers.

\subsection{Data Source}
\label{sec:data1}

The input dataset for Task~1 consists of documents in different formats and different levels of encoding quality and semantics:
\begin{itemize}
\item one HTML 4 index page linking to \emph{all} workshop proceedings volumes (\url{http://ceur-ws.org/}; invalid, somewhat messy but still uniformly structured)
\item the HTML tables of contents of selected volumes.
  They link to the individual workshop papers.
  Their format is largely uniform but has gained more structure and more semantics over time, while old volumes remained unchanged.
  Microformat annotations were introduced with volume 559 in 2010 and subsequently extended.
  RDFa (in addition to microformats) was introduced with volume 994 in 2013, but its use is optional, and therefore it has been used in less than 5~\% of all volumes since then.
  Valid HTML5 has been mandatory since volume 1059 in 2013; before, hardly any volume was completely valid.
\item the full text (PDF or PostScript) of the papers of \emph{some} volumes
\end{itemize}
Challenges in processing tables of contents include the lack of standards for marking up editors' affiliations, for separating multiple workshops in joint volumes, for referring to invited talks\footnote{Possible keywords include “invited paper”, “invited talk”, or “keynote”.}, or for linking to 
all-in-one proceedings PDFs.

The training and evaluation datasets, TD1 and ED2 were chosen to comprise a fair balance of different document formats.
To enable reasonable quality assessment, the training data already comprised all volumes of several workshop series, including, e.g., Linked Data on the Web, and for some conferences, such as WWW 2012, it comprised all those of its workshops that were published with CEUR-WS.org.
In the evaluation dataset ED2, some more workshop series and conferences were completed.
The datasets are available at \url{http://challenges.2014.eswc-conferences.org/index.php/SemPub/Task1}.

\begin{table}
\centering

\footnotesize

\caption{Task 1 Data Sources}
\renewcommand{\tabularxcolumn}[1]{>{\arraybackslash}m{#1}}
\newcolumntype{Y}{>{\centering\arraybackslash}X}
\newcolumntype{Z}{>{\arraybackslash}X}

\begin{tabularx}{\textwidth}{ >{\hsize=1\hsize}Z  >{\hsize=1\hsize}Y  >{\hsize=1\hsize}Y  }
\toprule

 & Training Dataset (TD1) & Evaluation Dataset (ED1) \\
\midrule
Proceedings volumes & 54 &  91 \\ 
… including metadata of & 689 papers &  1645 papers \\ 
Full text of & 46 papers & 88 papers \\
\midrule
Volumes using RDFa & 3 & 4 \\
… using microformats only & 48 & 71 \\
\bottomrule

\end{tabularx}

\label{table:Task1DataSourcesTable}
\end{table}

\subsection{Queries}
\label{sec:queries1}

The queries were roughly ordered by increasing difficulty.
The initial queries were basic ones to help the participants get started, whereas most queries from Q1.5 onward correspond to quality indicators discussed in Section~\ref{sec:objectives1}:
\begin{description}
\item[Q1.1] List the full names of all \textbf{editors of the proceedings of workshop} $W$.\footnote{We did not ask the participants to disambiguate names.
    In the relatively small set of CEUR-WS.org workshop editors and authors, names are rarely ambiguous.
    We believe that cases of ambiguity can easily be fixed manually.}
\item[Q1.2] Count the \textbf{number of papers in workshop} $W$.\footnote{Counting can be hard, as some workshops comprise multiple sessions, and as pre-microformat volumes do not always properly distinguish prefaces or all-in-one proceedings volumes from proper papers.}
\item[Q1.3] List the full names of all \textbf{authors who have \mbox{(co-)}au\-thored a paper} in workshop $W$.
\item[Q1.4] Compute the \textbf{average length of a paper} (in numbers of pages) in workshop $W$.\footnote{Page numbers in tables of contents are optional and do not always start from $1$.}
\item[Q1.5 (publication turnaround)] Find out whether the proceedings of workshop $W$ were published on CEUR-WS.org before the workshop took place.\footnote{For old proceedings volumes, the date of publication is not listed in the table of contents, but only in the main index.}
\item[Q1.6 (previous editions of a workshop)] Identify all editions that the workshop series titled $T$ has published with CEUR-WS.org.
\item[Q1.7 (chairs over the history of a workshop)] Identify the full names of those chairs of the workshop series titled $T$ that have so far been a chair in every edition of the workshop that was published with CEUR-WS.org.
\item[Q1.8 (all workshops of a conference)] Identify all CEUR-WS.org proceedings volumes in which papers of workshops of conference $C$ in year $Y$ were published.%
\footnote{Conferences may be abbreviated, e.g., as “CCCCYYYY” or “CCCC'YY”.}
\item[Q1.9] Identify those papers of workshop $W$ that were \textbf{(co-)authored by at least one chair} of the workshop.
\item[Q1.10] List the full names of all \textbf{authors of invited papers} in workshop $W$.
\item[Q1.11] Determine the \textbf{number of editions} that the workshop series titled $T$ has had, regardless of whether published with CEUR-WS.org.
\item[Q1.12 (change of workshop title)] Determine the title (without year) that workshop $W$ had in its first edition.\footnote{
Here%
, we assumed that the “see also” links in the main index always point to previous editions of a workshop.
    In reality, they sometimes point to closely related but different workshops.}
\item[Q1.13 (workshops that have died)] Of the workshops of conference $C$ in year $Y$, identify those that did not publish with CEUR-WS.org in the following year (and that therefore probably no longer took place).
\item[Q1.14 (papers of a workshop published jointly with others)] Identify the papers of the workshop titled $T$ (which was published in a joint volume $V$ with other workshops).
\item[Q1.15 (editors of one workshop published jointly with others)] List the full names of all editors of the proceedings of the workshop titled $T$ (which was published in a joint volume $V$ with other workshops).
\item[Q1.16] Of the workshops that had editions at conference $C$ both in year $Y$ and $Y+1$, identify the \textbf{workshop(s) with the biggest percentage of growth} in their number of papers.
\item[Q1.17 (change of conference affiliation)] Return the acronyms of those workshops of conference $C$ in year $Y$ whose previous edition was co-located with a different conference series.
\item[Q1.18 (change of workshop date)] Of the workshop series titled $T$, identify those editions that took place more than two months later/earlier than the previous edition that was published with CEUR-WS.org.
\item[Q1.19 (internationality of a workshop)] Identify all countries that the authors of all regular papers in workshop $W$ were from.
\item[Q1.20 (institutional diversity of a workshop)] Identify those papers in workshop $W$ that were (co-)authored by people from the same institution as one of the chairs (including papers by chairs).\footnote{Names of institutions require normalisation.
    English papers may, e.g., use the unofficial name “University of Bonn”.
  German papers often prefer the short “Universität Bonn” over the full official name Rheinische Friedrich-Wilhelms-Universität Bonn.}
\end{description}

Q1.5 (partly), Q1.12, Q1.13, Q1.16 and Q1.17 relied on the main index; Q1.19 and Q1.20 relied on the full-text PDF.

As Task~1 also aimed at producing linked data that we could eventually publish at CEUR-WS.org, the participants were additionally asked to follow a uniform URI scheme: \url{http://ceur-ws.org/Vol-NNN/} for volumes, and \url{http://ceur-ws.org/Vol-NNN/\#paperM} for a paper having the filename \url{paperM.pdf}.

\subsection{Accepted Submissions and Winners}
\label{sec:submissions1}

We received and accepted three submissions that met the requirements.

Ronzano et al.~\cite{RonzanoEtAl:SemPub14} solved the problem of annotating old proceedings volumes without any semantic markup by using the ones with microformat markup to train an automated annotation system that would retrofit microformat annotations to the old volumes.
The system consults dumps of DBLP and WikiCFP data to more reliably identify author names and conference titles and acronyms.
Using several external web services, the extracted data are enriched with further bibliographical information from BibSonomy and linked to the external datasets DBLP and DBpedia.
This submission won the award for the \emph{most innovative approach}.

Kolchin/Kozlov~\cite{KolKoz:SemPub14} defined templates for the typical structures of proceedings volumes, e.g., with or without RDFa or microformats.
These templates extract the required data from the HTML pages using XPath and regular expressions.
Countries (from authors' affiliations) are linked to DBpedia, using a SPARQL query against the DBpedia endpoint.
Despite (or because of?) its relative simplicity, this submission won the \emph{best precision/recall award}.

Dimou et al.~\cite{DimouEtAl:SemPub14} took the template approach a step further, implementing them in the declarative language RML
.
RML, which has so far been capable of mapping CSV, XML and JSON to RDF, was extended to process non-XML HTML input.
HTML elements with microformat annotations were accessed using CSS3 selectors.
This submission focused on those proceedings volumes that used microformats, using a separate RML mapping definition for each proceedings volume.
While sharing most code, many of them were manually adapted to the specific structures of certain volumes.

\begin{table}
\centering

\footnotesize

\caption{Task~1 evaluation results}

\begin{tabular}{p{.23\textwidth}p{.13\textwidth}p{.13\textwidth}p{.20\textwidth}p{.13\textwidth}p{.13\textwidth}}
\toprule
Authors & \textbf{Overall\newline average\newline precision} & \textbf{Overall\newline average\newline recall} & Queries\newline attempted & Average\newline precision\newline on these & Average\newline recall\newline on these \\
\midrule
Ronzano et al.~\cite{RonzanoEtAl:SemPub14}
& 0.335 & 0.313 & 1–12, 14, 15   & 0.478 & 0.447 \\ 
Kolchin/Kozlov~\cite{KolKoz:SemPub14}
& 0.707 & 0.639 & 1–20           & 0.707 & 0.639 \\ 
Dimou et al.~\cite{DimouEtAl:SemPub14}
& 0.138 & 0.092 & 1–9, 11–13, 16 & 0.212 & 0.142 \\ 
\bottomrule

\end{tabular}

\label{table:Task1Eval}
\end{table}

\subsection{Lessons Learnt}
\label{sec:lessons1}

From the perspective of running the challenge, Task~1 was successful, in that it led to three submissions, which were not only technically quite different, but whose performances could also be distinguished clearly – even when only taking into account the queries that the participants addressed at all (cf.\ Table~\ref{table:Task1Eval}).
Two solutions primarily consisted of code specific to this task~\cite{KolKoz:SemPub14,RonzanoEtAl:SemPub14}, whereas Dimou et al.\ wrote task-specific mappings in the otherwise generic RML language~\cite{DimouEtAl:SemPub14}.
While proving the versatility of RML, this approach had the drawback of neglecting all steps RML had not been designed for, such as cleaning up messy input.
Malformed literals remaining in the LOD stopped three queries from working.
While it seems striking that the most innovative approach did not perform best, note that poor precision/recall results are partly owed to the participants focusing on a subset of the datasets for lack of time.
For example, only one tool processed the main index and the full-text PDF~\cite{KolKoz:SemPub14}.
Therefore, and because of its good performance and its relatively few technical requirements, we are currently rolling it out at CEUR-WS.org to become part of the publication process.\footnote{For a proof of concept, request RDF/XML from \url{http://ceur-ws.org/Vol-1191/}.}
Only time will, however, tell the maintenance effort required for adapting to subsequent changes of the template for a proceedings volume table of contents.

\section{Task 2: Extraction and Characterisation of Citations}
\label{sec:task2}

\subsection{Motivation and Objectives}\label{Task2Motivation}

The importance of citations for the scientific community is undeniable: researchers cite works that investigated the same problem they are facing, or papers that proposed a similar (or contrasting) solution, 
and so on.

Citations are also being increasingly used to evaluate the impact of a given research, under the assumption that if a paper A cites another paper B then B had some impact on A. The impact factor is a measure of the quality of a journal obtained by averaging the
impacts of the papers published by that journal in a given time interval. Citations are also used to evaluate the quality of the work of single researchers or teams or even universities. Just think about the \textit{h}-index~\cite{hirsch2005index} and its diffusion.

Mere citation counting, however, is not enough to evaluate the quality of research. Citations are not all equal. Is it fair, 
e.g.%
, to give the same relevance to a citation of a paper that introduced a ground-breaking theory, compared to a citation of a paper that contains a lot of errors? Also, should self-citations have the same relevance as others%
? How should a citation be evaluated if grouped with many others in a generic list?

These
and other
questions inspired us in the design of Task 2: our objective was to investigate methods and tools to characterise citations \emph{automatically}, so that the tasks of linking, sharing and evaluating research through citations could be done in a more precise way. 
Participants were asked to process a set of XML-encoded research papers and to build an annotated network of citations. There are many different ways of annotating a citation, for instance by making explicit its type (is it a self-citation?), 
its structural information (where is it placed in the the citing paper?)
or even its function (why a paper is being cited?). 

Participants were asked to make such information explicit, so that it could be exploited to answer a given set of queries. They were free to use their own ontological model, provided that the given queries could be translated and run on their dataset. The queries are described in the following section, after presenting the dataset on which they were launched. 

\subsection{Data Source}\label{Task2DataSource}

The construction of the input datasets was driven by the idea of covering a wide spectrum of cases. We collected papers encoded in XML JATS\footnote{\url{http://dtd.nlm.nih.gov}}, a language for encoding journal articles derived from the NLM Archiving and Interchange DTD, and 
its TaxPub extension for taxonomic treatments.
The papers were selected from two sources:

\begin{itemize}
\item \emph{PubMedCentral Open Access Subset}\footnote{\url{http://www.ncbi.nlm.nih.gov/pmc/tools/openftlist/}}, a subset of the PubMedCentral full-text archive of biomedical and life sciences articles from different journals and publishers; some of these are freely available for redistribution and reuse. 

\item \emph{Pensoft Biodiversity Data Journal and ZooKeys archive}, open access archives of XML documents owned by Pensoft
\footnote{\url{http://www.pensoft.net}} and freely available for redistribution and reuse. Pensoft publishes scientific books 
on
natural history. Some years ago they launched their first open access journal (ZooKeys), implementing several innovations in digital publishing and dissemination. Recently, they launched the Biodiversity Data Journal (BDJ) and the associated Pensoft Writing Tool (PWT) as the first workflow that puts authoring, peer-review, publication, and dissemination into a single online collaborative platform.
\end{itemize}

The selected papers are structurally very different from each other. First of all, they use different elements to encode citations: some citations are organised in complete records, others are contained in mixed structures, others are completely unstructured and stored as plain text. Also, different elements are used to encode data about the authors (that in some cases are listed with their full names, or with their initials, or with different abbreviation styles) and about the publication types (stored as attributes or elements). Furthermore the papers use different content structures (this information is useful to identify the position of citations) and different forms to express acknowledgements (useful to extract data about grants and fundings). Finally, some citation sentences make explicit the reasons why a paper is being cited, but these sentences have different forms.

The datasets TD2 (training) and ED2 (evaluation) are available at 
\url{http://challenges.2014.eswc-conferences.org/index.php/SemPub/Task2}.
Table~\ref{table:Task2DataSourcesTable} reports some statistics about these datasets. TD2 is a subset of ED2, composed of \emph{randomly} chosen papers. The final evaluation was performed on a randomly chosen subset of ED2 too. To cover all queries and balance results, we clustered input papers around each query and selected some of them from each cluster. Each cluster was in fact composed of papers containing enough information to answer each query, and structuring that information in different ways.

\begin{table}[h!]
\centering

\footnotesize

\caption{Task 2 Data Sources}
\renewcommand{\tabularxcolumn}[1]{>{\arraybackslash}m{#1}}
\newcolumntype{Y}{>{\centering\arraybackslash}X}
\newcolumntype{Z}{>{\arraybackslash}X}

\begin{tabularx}{\textwidth}{ >{\hsize=1\hsize}Y  >{\hsize=1\hsize}Y  >{\hsize=1\hsize}Y  }
\toprule

 & Training Dataset (TD2) & Evaluation Dataset (ED2) \\
 \midrule

Papers & 150 &  400 \\

Journals & 15 &  70 \\
 
Citations & $>$ 10000 & $>$ 25000  \\

Citations per paper & $>$ 70 & $>$ 66 \\

Bibliographic references & 5419 &  16626 \\
\bottomrule

\end{tabularx}

\label{table:Task2DataSourcesTable}
\end{table}

\subsection{Queries}\label{Task2Queries}

The community has proposed many research quality indicators; the discussion about them is open.
We asked participants to extract some qualitative information about citations\todo{CL@ADI: I deleted the rest of this sentence as it is now redundant with the description of the evaluation procedure in section~\ref{sec:extraction-tasks}}
. 
The queries are not meant to be exhaustive but they were selected to provide a quite large spectrum of information. 
We selected 10 queries, covering different aspects of citations. The first two are very basic and are meant to check if the produced network of citations was complete and if the cited resources were classified correctly:

\begin{description}
\item [Q2.1] Identify \textbf{all papers cited} by the paper $X$
\item [Q2.2] Identify \textbf{all journal papers cited} by the paper $X$
\end{description}

In order to extract such information participants were basically asked to parse input files covering all possible cases: highly-structured citations, semi-structured and unstructured ones. 

The second group of queries was meant to check if the produced dataset contained enough information to identify authors and to find self-citations:

\begin{description}
\item [Q2.3] Identify \textbf{all authors cited by the author} whose surname is $X$
\item [Q2.4 (auto-citations)] Identify all papers cited by the paper $X$ and written by the same authors (or some of them)
\end{description}

The correct identification of the authors is tricky and opens complex issues of content normalisation and management of homonymity. A simplified approach was adopted for this task: participants were required to extract all information available in the input dataset and to normalise it by providing surname, first-name and first-name-initials. These data were all normalised in lowercase -- stripping spaces and punctuations -- in the final evaluation.

The following two queries covered the position and context of citations:

\begin{description}
\item[Q2.5] Identify all \textbf{papers cited multiple times} by the paper $X$
\item[Q2.6] Identify all \textbf{papers cited multiple times in the same paragraph} by the paper $X$
\end{description}

The idea behind these queries is that knowing the position of citations and their co-presence with others could give more information about their value (for instance, giving less relevance to multiple citations of the same work from the same paper, or even from the same paragraph). 

The last four queries required additional processing of the textual content of the papers. First, we added a query about grants and funding agencies:

\begin{description}
\item[Q2.7 (grants and funding agencies)] Identify the grant (or more than one) that supported the research presented in the paper $X$, along with the funding agency that funded it
\end{description}

The basic idea was to make such information explicit, so that it can be used to investigate how fundings were connected to, or even influenced, a given research. Information about fundings are examples of {\em research context data}. 
Many more contextual data could be extracted, for instance about the institutions involved in a research project 
or
partial works, etc. At this stage, we wanted to investigate some of these dimensions and to study how such structured data can be extracted from unstructured ones.

The last queries was meant to check how the tools characterised citations and to what extent they were able to identify the reasons why a work was cited:

\begin{description}
\item [Q2.8] Identify the \textbf{‘literature review’ section} of the paper $X$
\item [Q2.9 (using methods of a paper)] Identify all papers of which paper $X$ declares to use methodologies or theories
\item [Q2.10 (extending results of a paper)] Identify all papers of which paper $X$ declares to provide an extension of the results
\end{description}

Automatic characterisation of citation has previously been approached with CiTalO~\cite{DBLP:conf/esws/IorioNP13a}, a chain of tools based on Semantic Web technologies, and by machine learning~\cite{Teufel:2006:ACC:1610075.1610091}. In both of these cases, however, the agreement between human annotators and automatic processes was quite low.
Such a characterisation is actually an extremely difficult task for humans too: 
the experiments in~\cite{DBLP:conf/esws/CiancariniINPV14} confirmed that the opinions about citation functions are often disaligned, and rarely match among users. To avoid this problem we selected some simple cases and unambiguous forms of citations. Nonetheless these queries still ended up being too difficult and only a few 
were answered correctly %
(see below for details).

\subsection{Accepted Submissions and Winner}\label{Task2Submissions}

We received fewer submissions than expected. Some were incomplete and could not be considered for the evaluation. Eventually only one submission was completed and unfortunately we had to cancel the competition. 
Bertin and Atanassova~\cite{BerAta:SemPub14} presented a novel approach combining machine-learning techniques and rule-based transformation, which produced good results. 
We evaluated their tool as explained in Section~\ref{sec:extraction-tasks} even if they could not win an award.
They also provided us with a lot of feedback, useful to better shape such a task in the future.

\subsection{Lessons Learnt}\label{Task2Lessons}

The spirit of the challenge, and in particular of task 2, was to explore a large number of aspects in order to have a clearer picture of the state-of-the-art and to identify the most interesting and challenging issues.

In retrospect, this choice led us to defining a quite difficult task and we could have structured it in a slightly different way.
In fact the questions we were asking are logically divided in two groups: queries Q2.1---Q2.7 required participants to basically map data from XML to RDF, while the last three required additional processing on the content. These two blocks required different skills and some people were discouraged to participate as they only felt strong in one of them. We could have split them in two tasks: one on data conversion and publishing and another one on textual processing.

These two blocks also differ for the quality of the results. The proposed solution, in fact, showed that very good results can be achieved in producing \emph{semantically-annotated networks of citations}, that also include information about authors and about citation contexts. The automatic analysis of content and characterisation of citations, on the other hand, still has several open and fascinating issues to address.

\section{In-Use Task 3: Semantic Technologies in Improving Scientific Production}
\label{sec:task3}

\subsection{Motivation and Objectives}
\label{sec:objectives3}


The goal of task 3 was to investigate novel approaches to exploit 
semantic publications%
. We invited demos that showcased the potential of Semantic Web technology for enhancing and assessing the quality of scientific production. 
The task was open: participants could use their own datasets and were not required to connect to other tasks.
We followed the tradition of open challenges established in the community: 
in 2009 and 2011, e.g., Elsevier ran two Grand Challenges, for which challengers were asked to “1. improve the process/methods/results of creating, reviewing and editing scientific content; 2. interpret, visualize or connect the knowledge more effectively, and/or
3. provide tools/ideas for measuring the impact of these improvements”~\cite{Elsevier:GrandChallenge09} and to “improve the way scientific information is communicated and used”~\cite{Elsevier:EPC11}.
Task 3 is also similar to the open track of the ISWC Semantic Web Challenge\footnote{\url{http://challenge.semanticweb.org/2014/criteria.html}} (yearly since 2003) and the AI Mashup challenge\footnote{\url{http://aimashup.org}} (yearly since 2009), though focused on semantic publishing.

\subsection{Accepted Submissions}
\label{sec:submissions3}

We accepted 4 out of 5 submissions
. Each 
got 3 reviews and presented a sophisticated application for supporting researchers in their activities.

Linkitup~\cite{HoekstraEtAl:SemPub14} is a Web-based application for integrating research articles with semantic data retrieved from multiple heterogeneous sources. The platform enriches data available in existing repositories in many different ways: for instance, it finds terms, categories, entities (people, institutions, projects, etc.) related to a given work and shows them in an intuitive interface.
 
ROHub~\cite{PalmaEtAl:SemPub14} is a digital library for Research Objects (ROs) that supports their management, dissemination and preservation. ROs are defined as aggregations of scientific resources, not only papers but also experimental data, reports, slides, and so on. Particularly interesting is the support for the lifecycle of these objects, even in their drafting stage.

Rexplore~\cite{OsbMot:SemPub14} is a sophisticated Web-based platform for exploring and making sense of scientific data.
It integrates multiple sources and exploits data mining and semantic technologies to identify trends in research communities, to mine relations between researchers, to evaluate their performance, and to identify research trajectories. These data are shown in a rich interactive interface where users can search data, access them in personalised views and easily customise the overall dashboard by activating/deactivating the modules of the platform.

Atanassova and Bertin~\cite{AtaBer:SemPub14} presented an information retrieval system for enhancing publications by automatically identifying semantic relations between the components of these publications, for instance the methods, definitions and hypotheses. The system is based on a rule engine and offers a user-friendly interface that allows users to search, browse and filter results by using facets.

\subsection{Evaluation and Winners}
\label{sec:winners3}

In contrast to 
Tasks~1 and 2%
, the evaluation of Task~3 was carried out by an expert jury, taking into account the reviews and comments from the PC and applying five criteria:
\emph{potential impact} (the ability of the tool to make an impact on a large audience, with different background and expertise), 
\emph{originality} (its innovative nature compared to related work),
\emph{breakthrough} (to what extent the tool is groundbreaking and visionary, and opens new perspectives and challenges for researchers),
\emph{quality of the Demo} (clarity and usability of the demo), and
\emph{appropriateness for ESWC} (to what extent Semantic Web technologies play a prominent role in the tool). 
%
%
The jury finally decided to award Francesco Osborne and Enrico Motta for \emph{Understanding Research Dynamics}.
Both the robustness of the tool and its potential applications were appreciated: the ability to mine new unexpected information from existing data was considered a key success factor, together with the ability to integrate multiple resources, to identify research trends and to evaluate research results in an innovative way.

\section{Overall Lessons Learnt for Future Challenges}
\label{sec:over-less-learn}

The first lesson we learnt is that is challenging to define appealing tasks that bridge the gap between building up initial datasets and exploring possibilities for innovative semantic publishing.
As this first challenge has 
produced
linked data about the CEUR-WS.org workshops (currently being published) and subsets of the PMC Open Access and the PenSoft archives (not yet published, but reusable under an open license), we now have a foundation to build on.
Possible tasks for future challenges could focus on linking these initial datasets, each extracted from a single source, to \emph{further} relevant datasets, e.g., to link CEUR-WS.org workshops to co-located conferences in Springer's conference proceedings data~\cite{BrylEtAl:SePublica2014}, to identify the DBLP counterparts of articles and authors in the PMC and PenSoft datasets, or to link publications to related social websites such as SlideShare 
or Twitter.
Instead of a completely open task, one could call for applications that make innovative use of the data produced by the previous challenge.
Task suggestions from our participants addressed practical needs of researchers, such as finding high-profile venues for publishing a work, summarising publications, or helping early career researchers to find relevant papers
.

\paragraph{Acknowledgements}
We would like to thank our reviewers, judges and sponsors, whose names are mentioned in the preface of this overall proceedings volume, as well as our participants for their hard work, creative solutions and useful suggestions.
We would also like to thank Silvio Peroni for his suggestions on the overall challenge structure and, in particular, on the definition of task 2.

\printbibliography
\end{document}

